\documentclass{ws-ijmpd}
\usepackage[super,compress]{cite}
\usepackage{latexsym}
\usepackage{amstext}
\usepackage{graphicx}
\usepackage{subfigure}
\usepackage{mathrsfs}
\usepackage{dcolumn}
\usepackage{bm}
\usepackage{color}
\usepackage{enumerate}
\usepackage{soul}
\usepackage[normalem]{ulem}
\usepackage{verbatim}
\usepackage{ulem}
\usepackage{cancel}
\usepackage{etoolbox}

\date{\today}
\makeatletter
\gdef\@fpheader{}
\makeatother
\begin{document}
\markboth{Andr\'es Lizardo, Javier Chagoya, C. Ortiz}
{On phenomenological parametrizations for the luminosity distance
of gravitational waves}

\title{On phenomenological parametrizations for the luminosity distance
of gravitational waves}
\author{Andr\'es Lizardo, Javier Chagoya, C. Ortiz}
\address{Unidad Acad\'emica de F\'isica, Universidad Aut\'onoma de Zacatecas, Calzada Solidaridad esquina con Paseo a la Bufa S/N C.P. 98060, Zacatecas, M\'exico.}
\maketitle
\begin{abstract}
\noindent The propagation of gravitational waves offers new possibilities for testing the theory of gravity. Amongst these possibilities there is the luminosity distance of gravitational waves, $d_{gw}$. A few phenomenological parametrizations for this property have been proposed in the literature, but their generality is still under study. In this work we contribute to this study by confronting these parametrizations to the predictions of quadratic and degenerate higher order gravity. Furthermore, 
we propose a novel  parametrization that  unifies some of the existing proposals. 
We expect our findings to be relevant for future constraints on modified gravity based on the properties of standard sirens. 

\end{abstract}

\section{Introduction} \label{Int}

    
    
%
%
The direct detection of gravitational waves
(GW)~\cite{Abbott:2016blz,TheLIGOScientific:2017qsa} opened up new avenues for testing the theory of gravity. For instance, a difference between
the  propagation of  gravitational and electromagnetic waves over cosmological distances can be traced back to a modification of General Relativity (GR). Indeed, important constraints on Modified Gravity (MG) theories have been imposed from the difference on measurements of  the electromagnetic and gravitational waves produced by a binary neutron star inspiral, GW170817 \cite{TheLIGOScientific:2017qsa, Ezquiaga:2017ekz,Creminelli:2017sry,Baker:2017hug,Copeland:2018yuh,Kase:2018aps,Bordin:2020fww}.  However, an electromagnetic counterpart is not absolutely necessary in order to use GWs to derive constraints on MG, for instance, strong constraints that do not rely on an electromagnetic counterpart have been derived in, e.g., \citen{Finke:2021aom,Mancarella:2021ecn,Leyde:2022orh}.

Testing the theory of gravity with GW at cosmological scales is particularly relevant;  it is precisely at these scales where many theories of MG purposely deviate from GR in order to alleviate the cosmological constant problem~\cite{Weinberg:1988cp,Padilla:2015aaa}. By doing so, these theories introduce new dynamics that might affect the behaviour of tensor perturbations (e.g.~[\citen{Kobayashi:2011nu}]). Therefore, MG theories are subject to GW constraints,
even if they are endowed with screening mechanisms~\cite{Brax:2013ida, Joyce:2014kja} that allow them to recover GR predictions at Solar System scales.

Since the first few years after the introduction of GR,  interest in MG has been continuously sourced mainly by theoretical arguments,  (e.g. [\citen{Clifton:2011jh,Koyama:2015vza}] for reviews). This has led to
a plethora of models that address issues such as the 
nature of dark energy and dark matter or the fate of GR at high energy scales.
Although the motivations for these models  emerge from diverse scenarios, it turns out that some of their
predictions for physical observables can be characterised by a few parameters. This feature is exploited, for instance, in the study of post-Newtonian solutions~\cite{Will:2014kxa}, cosmological evolution~\cite{Mukhanov:1990me,Baker:2012zs,Gleyzes:2013ooa}, and inflationary mechanisms~\cite{Cheung:2007st}. These 
parametrized approaches provide a 
useful tool for testing generic features of MG theories. In particular, some parametrizations have
been proposed for the study of modifications to the friction term in the gravitational wave equation~\cite{Belgacem:2018lbp, Belgacem:2019pkk, LagosSirens,Matos:2021qne}.

In this work we analyse the parametrizations proposed by Belgacem et. al.~\cite{Belgacem:2019pkk}. We focus on their ability to characterise the predictions of a model within Degenerate Higher Order Scalar-tensor Theories (DHOST) of gravity~ \cite{langlois2018degenerate} and a model of quadratic gravity. We remark their strengths and weaknesses and then propose a novel parametrization that fits a wider range of models of modified gravity.

{DHOST gravity is a class of theories that has received attention} in the past few years. These theories are a generalisation of  Hordenski gravity which comes up as a way to avoid the Ostrogradsky instability \cite{woodard2015theorem}. DHOST modifies the Lagrangian of GR by the addition of higher order terms of a scalar field, and it focuses on the degeneracy of their Lagrangian rather than the order of their equations of motion \cite{langlois2018degenerate}. By properties demanded on the scalar field, DHOST is a ghost-free theory, and it has only 3 degrees of freedom (two for the metric and one in the scalar field). Also, due to its generality, it admits several cosmological solutions \cite{2016dhostcubic,2017effectivedescription}.

On the other hand, we also study a model of quadratic gravity known as Einstein-scalar-Gauss-Bonnet (EsGB), a theory that
modifies GR by the addition of 
 the Gauss-Bonnet invariant directly coupled
to a scalar field. In four dimensions, the Gauss-Bonnet term is
proportional to the Euler topological term. Thus, by itself, it does not
modify the classical equations of motion of GR. However, once it is coupled to a scalar field, it becomes dynamical. In the late 80's it was found that this type of 
coupling appears as a second order curvature correction in heterotic string theory~\cite{Callan:1985ia,Gross:1986mw,Metsaev:1987zx}, therefore motivating investigations on the physical significance of a scalar-Gauss-Bonnet theory. Along these lines it was shown  \cite{Rizos:1993rt,Antoniadis:1993jc, Kanti:1998jd} that the theory admits singularity-free solutions
for a spatially-flat Friedman-Robertson-Walker (FRW) metric, prompting a sustained interest in the study of inflationary, late-time, and astrophysical solutions
to EsGB and related models~\cite{Koivisto:2006ai,Sanyal:2006wi,Kanti:2015pda,Heydari-Fard:2016nlj,Sberna:2017xqv, Odintsov:2018zhw, Odintsov:2019clh,Odintsov:2020sqy,Carson:2020ter}.
Another relevant property of EsGB is that it is 
free of Ostrogradsky instabilities, unlike generic higher order curvature theories which introduce derivatives
of order higher than two in the equations of motion (see [\citen{Woodard:2006nt}] for a review of Ostrogradsky theorem). 
Finally, recent investigations on astrophysical solutions have found that future observations of waveforms of the inspiral phase of binary black hole systems have the potential to constrain EsGB gravity~\cite{Carson:2020ter}, while black hole shadows could constrain the theory if the uncertainty in
ring diameter measurements decreases from a relative error of 6\% to about 1\%~\cite{Cunha:2016wzk}. In view of the above, it is interesting to continue the investigation of phenomenological signatures of EsGB. 



This work is organised as follows. In Sec.~\ref{sec:linwave}
we describe the equation that governs gravitational wave
propagation, its modifications in MG models,
and the parametrizations that attempt to
capture these modifications. In Sec.~\ref{sec:ap}
we propose a novel parametrization that is valid for a wider range of models. In
Sec.~\ref{sec:dhost} and Sec.~\ref{sec:EGB} we confront these parametrizations to DHOST and EsGB gravity models. 
In Sec.~\ref{sec:results}
we analyse the results of such tests.  Sec.~\ref{sec:disc} is devoted to discussion and concluding
remarks.

\section{Luminosity distance of gravitational waves}
\label{sec:linwave}
In GR, linearised tensor perturbations on a Friedmann-Robertson-Walker background are described
by the equation
\begin{equation}
    \tilde h''_A + 2 \mathcal H \tilde h'_A + c_{gw}^2 k^2 \tilde h_A = \Pi_A\,, \label{eq:linpertgr}
\end{equation}
where  $\tilde h_A(\eta, \mathbf k) = (\tilde h_+, \tilde h_\times)$ are the Fourier modes of the GW amplitude, $k$ is the wavenumber, $\eta$ stands for conformal time, primes
denote derivatives with respect to $\eta$, $\mathbf k$ is the wave vector, $a(\eta)$ the scale factor,  $\mathcal H = a'/a$ is the Hubble parameter, and $\Pi_A$ is a source term. {In GR the speed of propagation of GWs is equal to the speed of light, $c_{gw}=c$. Recent observations place bounds on this
quantity at the level $|c_{gw} -c|/c < 10^{-15}$~\cite{Monitor:2017mdv}.} The first derivative in eq.~\eqref{eq:linpertgr} corresponds to a friction term
whose coefficient depends on the change of the scale factor. 
{From now on, we set $c=1$}.

Generic modifications of gravity affect each term in eq.~\eqref{eq:linpertgr}.  From now on, we focus on
theories where the source term $\Pi_A$ is not present. We  also drop modifications to the third term, 
i.e., to the speed of propagation of GW. This is completely valid in theories that predict $c_{gw}=1$, such as a reduced set of (beyond) Horndeski theories that includes quintessence, Brans-Dicke, $f(R)$ and other popular models~\cite{Ezquiaga:2017ekz}. Theories that do not generically predict 
$c_{gw}=1$ are disfavoured, however, even within such theories, it is possible to construct models that are compatible with $c_{gw}\approx 1$ under special circumstances~\cite{Odintsov:2020sqy, Bordin:2020fww}. {Also, it is important to notice that this constrain is strictly valid only in the local Universe, thus, for $z\gtrsim 0.1$ one could consider the full Horndeski
theory, as done in~\cite{DAngostinoSirens}}. Here, we consider a modified wave equations with $c_{gw}=1$, 
\begin{equation}
    h_A'' + 2 \mathcal H 
    \left[ 
    1-\delta(\eta)
    \right]h'_A + k^2 h_A = 0\,, \label{eq:mglinpert}
\end{equation}
where $\delta$ represents the modifications to the friction term arising from MG. {Due to the couplings between additional fields and curvature that appear in several modified gravity theories, $\delta(\eta)$ may depend on 
the background solutions of these fields.} The function $\delta(\eta)$ affects the amplitude
of GW, i.e., the amount by which the amplitude of a wave decreases from the time of emission to the time of observation
depends not only on the scale factor, but also on the function
$\delta(\eta)$. This leads to the notion of 
\textit{gravitational wave luminosity distance}, $d_L^{gw}$~\cite{Belgacem:2018lbp}. The distance  $d_L^{gw}$ is related to
the usual electromagnetic luminosity distance $d_L^{em}$, by
\begin{equation}
    d_L^{gw}(z) = d_L^{em}(z)\exp\left\{-\int_0^z \frac{dz'}{1+z'}\delta(z') \right\}\,, \label{eq:lumdis}
\end{equation}
where all functions are expressed in terms of redshift $(z)$
instead of time, and the normalisation $a(z=0) = 1$ is chosen without loss of generality. {From now on we use $\delta(z)$ and $\delta$ 
interchangeably.}
The difference between $d_L^{gw}$ and $d_L^{em}$ implies that measurements of luminosity distance performed with coalescing compact binaries  are different from measurements with electromagnetic probes, thus providing a possible test for deviations of GR, as explored, e.g.,  in~\citen{Escamilla-Rivera:2021boq,Calcagni:2019kzo,Fanizza:2020hat,Belgacem:2019pkk,Belgacem:2018lbp}. 

\subsection{Phenomenological Parametrizations}

Since an observation of $d_L^{gw}/d_L^{em}\neq 1$ does not necessarily point to a specific model of modified gravity, it
is convenient to think of {$\delta$} in terms of parametrizations that incorporate features expected in
generic theories of MG. For instance, if MG is designed
to solve the cosmological constant problem it would be
ideal for it to affect only late time cosmology, this implies that contributions of {$\delta$} to linearised perturbations should
vanish at high redshifts. On the other hand, at small
redshifts the function
\begin{equation}
    \Xi(z) \equiv  \frac{d_L^{gw}(z)}{d_L^{em}(z)},\label{eq:Xi}
\end{equation}
is expected to behave as $\Xi(z\ll 1)=1$, since at those values of $z$ the effects of modified 
GW propagation do not have enough time to accumulate 
into an observable quantity. Notice that the relation 
between $\Xi(z)$ and {$\delta$} can be written as
\begin{equation}
    \delta(z) = -\frac{d\ln\Xi(z)}{d\ln(1+z)}\,, 
\end{equation}
 thus $\Xi(z)$ approaches a constant value for large $z$. 
 
With the considerations above in mind, Belgacem et al.~\cite{Belgacem:2018lbp} propose
a parametrization for {$\delta$} that interpolates between
the expected behaviours at large and small $z$,
\begin{equation}
    \delta_{I}(z) = \frac{n(1-\Xi_0)}{1-\Xi_0 + \Xi_0 (1+z)^n}\, , \label{eq:param1}
\end{equation}
where $n$ and $\Xi_0$ are free parameters, both assumed positive. It is straightforward to
verify that $\delta_I(z\gg 1)= 0$ and
$\delta_I(z\ll 1) = n(1-\Xi_0)$.
The quantity $\delta(z)$ is of interest since it 
represents the modification to the GW equation due to MG; however, since the observables are $d_L^{gw}$ and $d_L^{em}$, it is also convenient
to express the parametrization in terms of $\Xi(z)$. Corresponding to eq.~\eqref{eq:param1} we have
\begin{equation}
    \Xi_{I}(z) = \Xi_0 + \frac{1-\Xi_0}{ (1+z)^n}\,.\label{eq:param1xi}
\end{equation}
with $\Xi_I(z\gg 1) = \Xi_0$ and $\Xi_I(z\ll 1) = 1$. Indeed, at small $z$, $\Xi_I(z)$ deviates from unit by terms of order $z$. 

Since $\Xi(z)$ involves
an integration over $z$, it is less sensitive
than {$\delta$} to details of GW propagation -- for instance, a peak in {$\delta$} might not be reflected in $\Xi(z)$. In the spirit of parametrizing generic deviations from GR, this is a positive aspect. In principle, either eq.~\eqref{eq:param1} or eq.~\eqref{eq:param1xi} is to be fitted against observational data. However, first, it has to be checked that these
parametrizations do indeed describe  effects induced on GW propagation by generic MG models. Some steps in this direction have been taken, for instance, in Belgacem et al.~\cite{Belgacem:2019pkk}, where
a set of parametrizations including eqs.~\eqref{eq:param1}, \eqref{eq:param1xi} were fitted to the explicit predictions of some well motivated MG models, finding that, while in most cases eqs.~\eqref{eq:param1}, \eqref{eq:param1xi} provide a good fit,
exploring alternative parametrizations is important in order 
to account for more complicated, but still well motivated,
physical predictions of MG. Specifically, the following alternative parametrizations for {$\delta$} are explored in the same reference,
\begin{eqnarray}
    {} \delta_{II}(z) = \delta_I(z)
    + \frac{n(1-\Xi_0)}{(1+z)^n} - \frac{2n(1-\Xi_0)}{(1+z)^{2n}},\label{eq:param2} \\
    {} \delta_{III}(z) = \frac{n(1-\Xi_0)(1+z)z^{n-1}}{1+\Xi_0(e^{z^n} - 1)}\label{eq:param3}.
\end{eqnarray}
Equivalently, the parametrizations for
$\Xi(z)$  are
\begin{eqnarray}
    {} \Xi_{II}(z) &=& \exp
    \left\{ 
    -\frac{(1-\Xi_0)\left[ 1 - (1+z)^n\right]}{(1+z)^{2n}}\right\}
    \left[ 
    \Xi_0 + (1-\Xi_0)(1+z)^{-n}
    \right], \\
     \Xi_{III}(z) &=& \Xi_0 + (1-\Xi_0) e^{-z^n}.\label{eq:paramxi3}
\end{eqnarray}

It is important to notice that the parametrization
$\delta_I$ is accurate for models where {$\delta$} grows near $z=0$, as expected for theories that modify gravity at late times. On the other hand, the second and third parametrizations ($\delta_{II}$, $\delta_{III}$) are designed for models where {$\delta$}$\sim 0$  near $z=0$, this is known to happen in a subset of DHOST
gravity with a fixed de Sitter point and a 
$\delta(z)$ proportional
to the time derivative of the Hubble factor\cite{Belgacem:2019pkk}.  Therefore, it is relevant to design a 
parametrizations that automatically selects whether the model prefers {$\delta\to 0$} or not near $z=0$.  This is done in the next section.

 \section{A novel parametrization}
 \label{sec:ap}
In this section we present a more general parametrization that interpolates between the behaviours of $\delta_I$ and $\delta_{II}$, thus being able to fit models that evolve towards a de Sitter fixed point as well as models that do not, {at the price of introducing an additional parameter.}  The construction is motivated by the pattern observed in the parametrizations $\delta_{I}$ and $\delta_{II}$ (eqs. \eqref{eq:param1}, \eqref{eq:param2}). Specifically, we see that $\delta_{II}$ modifies $\delta_I$ by the addition of two terms that suggest a power series in $1/(1+z)$. The function $\delta_{II}$ is designed to fit models that approach a de Sitter fixed point as $z = 0$, while $\delta_I$ does not follow this requirement. Since $\Xi_0$ is assumed to be positive, we notice that
$\delta_{II}$ is driven to zero near $z=0$ only by its last term. Therefore, we generalise $\delta_{II}$ by replacing its second and third terms with two 
power series, each weighted by an independent parameter, such that the first series adds up to the magnitude of the first term (i.e. to $\delta_I$) and the second one can reduce this magnitude  if necessary. We propose a parametrization $\delta_X$ given by
\begin{eqnarray}
    {} \delta_{X}(z) & = & \frac{n(1-\Xi_0)}{1-\Xi_0+\Xi_0(1+z)^n}+ \frac{n(1-\Xi_0)}{(1+z)^n} - \frac{2r(1-\Xi_0)}{(1+z)^{2r}} + \frac{3n(1-\Xi_0)}{(1+z)^{3n}} - \dots 
\end{eqnarray}
Assuming an infinite number of terms, this parametrization can be rewritten as 
\begin{equation}
    {} \delta_{X}(z)  =  \delta_I(z) + \frac{n(1-\Xi_0)}{(1+z)^n} \left[\sum_{i=1}^\infty \frac{2i-1}{((1+z)^{2n})^{i-1}}\right] - (1-\Xi_0)\left[\sum_{i=1}^\infty \frac{2ir}{(1+z)^{2ir}}\right]\, ,
\end{equation}
the sum in the first square bracket is further separated into two sums, one with coefficient $2i$ and the other with coefficient $-1$. The latter is immediately recognised as a geometric series, while the former is the derivative of the geometric series. Similarly, the sum in the second square bracket  is also the derivative of a geometric series. For $z>0$ these geometric series are convergent. If we want to include $z=0$, regularization is needed. For simplicity we define $p = 1+z$, and after a few simplifications $\delta(z)$
can be rewritten as follows,


\begin{equation}
{} \delta_X(p) = (1-\Xi_0)\left(\frac{n\textit{p}^n(1+\textit{p}^{2n})}{(\textit{p}^{2n}-1)^2} - \frac{2r\textit{p}^{2r}}{(\textit{p}^{2r}-1)^2} + \frac{n}{1+\Xi_0(\textit{p}^n-1)}\right). \label{eq:delx}
\end{equation}

The corresponding parametrization for $\Xi_X(z)$ is, after regularization at $z=0$, given by
\begin{equation}
{} \Xi_X(p) =    \Xi_I \exp \left[ \left(\Xi_0-1\right) \left(\frac{\textit{p}+1}{4n (\textit{p}-1)}-\frac{\textit{p}+1}{4r (\textit{p}-1)} -\frac{ \textit{p}^n}{\textit{p}^{2 n}-1}+\frac{1}{\textit{p}^{2 r}-1}+\frac12\right)\right]. \label{eq:xix}
\end{equation}
This $\Xi_X$ satisfies $\Xi_X(z = 0) =1$, and near this region it differs from unit by terms of order $z$. On the other hand, for large redshifts, $\Xi_X(z\gg1) = \Xi_{X0} $, where the constant $\Xi_{X0}$ is determined by $\Xi_0, n$ and $r$. It is worth noting that the large $z$ behaviour of $\Xi_X(z)$ is determined by the powers $n$ and $r$, while at low $z$ the behaviours is linear. This characteristic is shared with $\Xi_{I}$, $\Xi_{II}$ and $\Xi_{III}$, and contributes to the generality of the parametrizations.  {We emphasize that 
$\Xi(z)$ is more directly related to observables than $\delta(z)$, which is derived from the former. {It is then important that $\Xi_X$ has the properties mentioned above, but it is also important to analyse whether the addition of a third parameter has a significant effect. As we
will see below, due to the cumulative nature of $\Xi(z)$ this is not the case.  }{In addition, one has to analyse the effect 
of the third parameter 
on the goodness of fit weighted by degrees of freedom. We perform this analysis in the following sections using reduced $\chi^2$ and adjusted $\mathcal R^2$.} On the other hand, $\delta_X$ divides into two classes depending  on the values of $n$ and $r$: for $n\leq r$, $\delta_X$ is a monotonic function of $z$, while for $n>r$ it is not. Thus, $\delta_X$ may give us more information than $\Xi_X$ about the underlying theory causing the modifications to the friction term in the gravitational wave equation.}
In the next sections we derive theoretical predictions for $\delta(z)$ in DHOST and quadratic gravity, and later on we confront the parametrizations presented so far to these predictions.
\section{Gravitational wave luminosity distance in a DHOST gravity scenario}
\label{sec:dhost}
Degenerate Higher Order Scalar-Tensor theories of gravity are the most general 
scalar-tensor theories that propagate 3 degrees of freedom, two from the metric and one from the scalar field. Unlike Horndeski gravity, in DHOST, the equations of motion can be of order higher than two, due to the existence of constraints that avoid the propagation of additional degrees of freedom. A large number of theories can be constructed within DHOST~\cite{2016dhostcubic,2017effectivedescription}. A restricted set, obtained by
requiring that the speed of gravitational waves equals the speed of light, is 
described by the action
\begin{align*}
S = \int d^4x \sqrt{-g}
&\left[
K + G_3\Box\phi + f_2 R + A_3\Box\phi \phi^\mu \phi_{\mu\nu}
\phi^\nu + A_4 \phi^\mu \phi_{\mu\rho}\phi^{\rho\nu} \phi_\nu \right. \nonumber \\
&\left. +
A_5 (\phi^\mu \phi_{\mu \nu} \phi^\nu)^2
\right],
\end{align*}
where $\phi$ is a scalar field, $\phi_\mu\equiv\nabla_\mu\phi$, and $K, G_3, f_2, A_3, A_4, A_5$ are free functions of $\phi$ and $X$, with $X\equiv\nabla_\mu\phi\nabla^\mu\phi$. In order to avoid the propagation of additional degrees of freedom, these functions have to be related by
\begin{eqnarray}
 A_4 & = &-\frac{1}{8 f_2} \left[
8 A_3 f_2 - 48 f_{2X}^2 - 8 A_3 f_{2X} X
+ A_3^2 X^2\right], \\
 A_5 & =& \frac{A_3}{2 f_2} \left(
4 f_{2X} + A_3 X\right),
\end{eqnarray}
A particular case of the restricted DHOST action with the choices
\begin{equation}
    K = c_2 X, \ \ G_3 = \frac{c_3}{\Lambda_3^3}X, \ \ 
    f_2 = \frac{M_{Pl}^2}{2} + \frac{c_4}{\Lambda_3^6}X^2, \ \ 
    A_3 = -\frac{8c_4}{\Lambda_3^6} - 
    \frac{\beta}{\Lambda_3^6},
\end{equation}
where $M_{Pl}$ is the Planck mass,
$c_i$ are constants, and $\Lambda_3$ is 
a coupling scale, has been  shown to admit tracker solutions $\dot\phi\propto H^{-1}$ both at the matter domination and
de Sitter expansion eras~\cite{Crisostomi:2017pjs, Crisostomi:2018bsp}, with different proportionality constants at each era. For this model, the modification to the
friction term in the GW equation is given
by~\cite{Belgacem:2019pkk},
\begin{equation}
    \delta(z) = -\frac{1}{H} \frac{4 c_4 \dot\phi^3 \ddot\phi}{1+2c_4\dot\phi^4},
    \label{eq:ddhost}
\end{equation}
which implies {$\delta\to 0$} at the de Sitter fixed point, where $\dot\phi$ takes a constant value.
Let us now show that $\delta_X$ is compatible with Eq.~\eqref{eq:ddhost}.
For simplicity, we compare to a $\delta_{II}$ that fits {$\delta$}
for a particular solution of DHOST~\cite{Belgacem:2019pkk}.


\begin{figure}
    \includegraphics[width=0.49\textwidth]{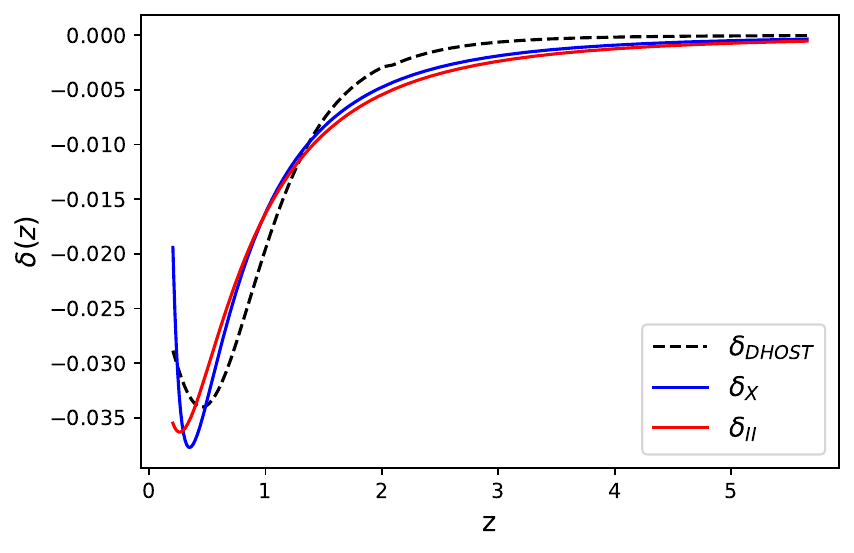}
    \includegraphics[width=0.49\textwidth]{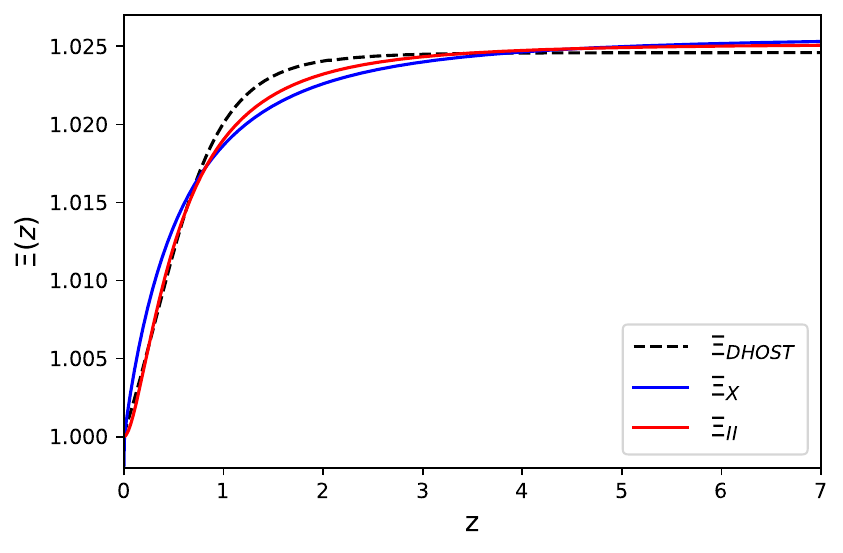}
    \caption{Parametrizations $\delta_{II}$ and $\delta_X$ (left), with their respective $\Xi(z)$ (right) compared to a prediction of a  DHOST gravity  model~\cite{Crisostomi:2018bsp}.}
    \label{fig:dhostdata}
\end{figure}

\begin{figure}
    \centering
    \includegraphics[width=0.45\textwidth]{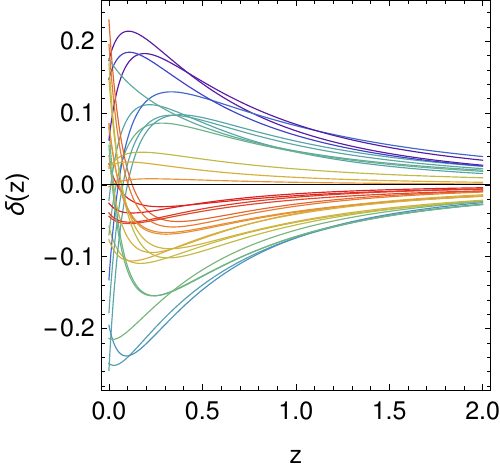} \ \ 
    \includegraphics[width=0.45\textwidth]{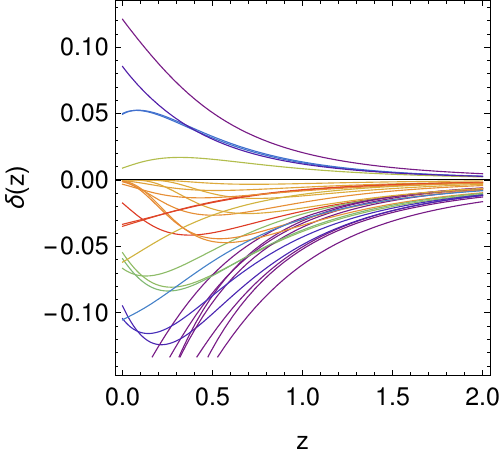}
    \caption{Left: profiles for $\delta_X$ with parameters given by random variations around the best fit to the DHOST model of Fig.~\ref{fig:dhostdata}. Right: profiles for $\exp[-(z+c)^{-2}]\delta_X$, again with random parameters around the best fit to the DHOST model of Fig.~\ref{fig:dhostdata}. In both cases, the variations around the best fit are in the ranges $\pm0.5$ for $n$ and $r$, and $\pm 0.2$ for $\Xi_0$. For the right panel, the range of variation for $c$  is such that $0<c<1.5$.} 
    \label{fig:deltaxgaussvsnotgauss}
\end{figure}

 In Fig.~\ref{fig:dhostdata} we confirm that $\delta_X$ and $\Xi_X$
are  able to approximate the type of modifications to the friction term and
to the luminosity distance that appear in DHOST theories with a fixed de Sitter point and a tracker cosmological solution. {The fits provided by $\delta_X$ are better than those provided by $\delta_{II}$ according both to $\chi^2$ and adjusted $\mathcal R^2$ estimators (Table~\ref{tab:x2dhost}); however, these estimators have to be taken with some care, the former since we are not using real data, and the latter because our regression is non-linear.  }
\begin{table}
\tbl{{Statistical estimators for the parametrizations $II$ and $X$ adjusted to the predictions of a model of DHOST gravity.}}
{\begin{tabular}{@{}ccccc}
       & \multicolumn{4}{c}{DHOST data} \\ 
\cline{1-5} 
        & $\delta_{II}$ & $\delta_X$  & $\Xi_{II}$ & $\Xi_X$\\ \colrule
        $\chi^2$ & $0.3744$ & $0.2436$ & $0.0919$ & $0.1394$ \\  \colrule
        $\chi^2_{red}$ & $0.000734$ & $0.000478$ & $0.00018$ & $0.000273$ \\  \colrule
        $\mathcal{R}^2$ & $0.9547$ & $0.9705$ & $0.9838$ & $0.9754$\\ \colrule
        $\mathcal{R}^2_{adj}$ & $0.9112$ & $0.9416$ & $0.9678$ &  $0.9514$\\
        \botrule
    \end{tabular}\label{tab:x2dhost}
}
\end{table}
By further analysing $\delta_X$, it is found that the peak shown in Fig.~\ref{fig:dhostdata} can be shifted around by changing the parameters $(n,r,\Xi_0)$, however the magnitude of its slope always gets larger as $z$ gets smaller, i.e., the peak is asymmetric. This is a limitation of $\delta_X$. One possible way around is to weight $\delta_X$ by a factor of
the form $\exp[-(z+c)^{-2}]$, with $c>0$, this introduces a $z$-dependent damping that smooths out the asymmetry of the peak while
also allowing it to be less extended, thus making it more similar to the predictions for DHOST gravity reported in~\cite{Belgacem:2019pkk}. {Figure~\ref{fig:deltaxgaussvsnotgauss} shows an illustration of this idea.}

\begin{figure}
    \centering
    \includegraphics[width=0.7\textwidth]{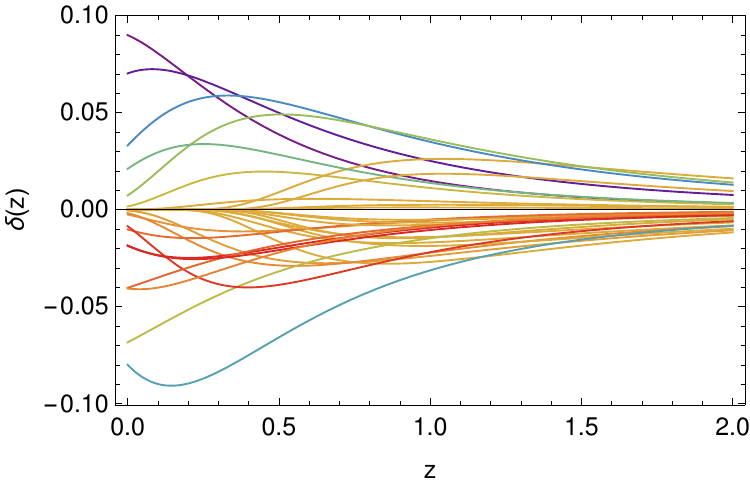}
    \caption{Profiles for $\delta_{Iexp}$ with random parameters in the ranges $n\in[2.5,4.5]$, $r\in[0.8,1.2]$,  $c\in[0,1]$.}
    \label{fig:expcorrection}
\end{figure}

{The same argument can be made about parametrizations I, II and III. This becomes more interesting for parametrization I, since without the exponential modification it was not able to fit models with $|\delta(z)|$ decreasing near $z=0$, while with this modification one could keep a simple functional form for $\delta$,
\begin{equation}
    \delta_{Iexp} = \exp[-(z+c)^{-2}]\delta_I,
\end{equation}
with only 3 parameters, $(n, c, \Xi_0)$, and get  fits appropriated to different classes of models. Fig.~\ref{fig:expcorrection} shows some of the profiles that can be obtained with $ \delta_{Iexp}$.} As in Fig.~\ref{fig:deltaxgaussvsnotgauss}, we see that the exponential factor introduces enough freedom to shift the peak of $\delta(z)$. 
The preceding discussion highlights the fact that there is still work to do on parametrizations that are general enough
without introducing a large number of parameters. On the other hand, $\Xi(z)$ -- which is more directly related to observations -- presents a simpler form. As the right panel of Fig.~\ref{fig:dhostdata}
shows its theoretical profile is similar to the ones obtained in models 
without a de Sitter fixed point, and the parametrization $\Xi_X$
is appropriate for this type of profiles, as one can conclude from 
Fig.\ref{fig:dhostdata}.

\section{Gravitational wave luminosity distance in Einstein-scalar-Gauss-Bonnet}\label{sec:EGB}
Einstein-scalar-Gauss-Bonnet (EsGB) gravity is a MG theory that adds higher order curvature terms to GR. Unlike other higher order curvature corrections, EsGB is free of Ostrogradsky instabilities (see [\citen{Woodard:2006nt}] for a review of Ostrogradsky theorem). Indeed, in four dimensions EsGB is the only  ghost-free theory with quadratic curvature terms.
{  Being a ghost-free scalar-tensor theory, EsGB is included in Horndeski's theory -- the most general scalar-tensor theory in 4 dimensions with second 
order equations of motion~\cite{Horndeski:1974wa, Deffayet:2009wt}. The identification is not trivial and is performed by studying the equations of motion \cite{Kobayashi:2011nu}. As is the case for the general Horndeski Lagrangian, EsGB is in general incompatible with the observational constraints on the speed of gravitational waves~\cite{TheLIGOScientific:2017qsa}; however, it was found
that compatibility can be achieved by imposing a 
differential equation on the scalar coupling function \cite{Odintsov:2019clh,Odintsov:2020sqy}. } Although we shall focus on this theory, it is worth mentioning that further generalisations where
the Gauss-Bonnet $\mathcal G$ term is replaced by a function
$f(\mathcal G)$ have been studied in the literature, and it has been shown that their ghosts can be removed at the level of the equations of motion by the addition of constraints via Lagrange multipliers~\cite{Nojiri:2018ouv}. In this work we consider an action
\begin{equation}
    S = \int d^4 x \sqrt{-g} 
    \left[ \frac{M_{Pl}^2}{2} R - X - V(\phi)- \xi(\phi)\mathcal G \right]\,,\label{eq:actionesgb}
\end{equation}
where the Gauss-Bonnet term is given in terms of the curvature tensor $R^{\mu\nu\alpha\beta}$ and its contractions as $\mathcal G = R^2 - 4 R_{\mu\nu}R^{\mu\nu}
+ R_{\mu\nu\alpha\beta}R^{\mu\nu\alpha\beta}$, $\xi(\phi)$ is the coupling between the scalar field $\phi$ and $\mathcal G$, 
$X = \partial_\mu\phi \partial^\mu\phi/2$, $V(\xi)$ is the scalar field potential, and $M_{Pl}$ is the 
Planck mass. It is worth mentioning that EsGB, eq.~\eqref{eq:actionesgb}, is part of Horndeski gravity~\cite{Kobayashi:2011nu}.
The potential $V(\phi)$  does not affect the friction term of the
gravitational wave equation, thus it will be neglected when analysing the gravitational wave luminosity distance (for a complete description of how each term in Horndeski gravity affects the GW equation see, e.g., [\citen{Arai:2017hxj}]); however, together
with the coupling  $\xi(\phi)$, it is relevant in determining the
background metric on which we study GW propagation. Different choices for the coupling and potential functions have been explored in the literature\cite{Rizos:1993rt,Soda:1998tr,Kanti:1998jd,Kanti:2015pda,Heydari-Fard:2016nlj,Odintsov:2019clh}, initially motivated by the existence of non-singular cosmological solutions. Shortly  after, it was realised that the theory has applications to the inflationary and late time acceleration eras. Furthermore, it has been shown that
there is a sector of the theory that predicts gravitational waves that propagate at the speed of light. As a sample of the possible
cosmological solutions to EsGB gravity, we work with the 
following models:
\begin{description}
\item[Model A] Based on a model that describes de Sitter expansion
with an exponential coupling function $\xi(\phi)\sim e^\phi$ and a scalar field that depends logarithmically on the scale factor, $\phi \sim \ln a(t)$, with a certain relation between the proportionality constants appearing in these functions\cite{Heydari-Fard:2016nlj}. In the spirit
of testing the validity of the $\delta$-parametrizations instead of  testing the validity of the model using the precise form of the solutions  \cite{Heydari-Fard:2016nlj}, we rather assume
 
\begin{eqnarray}
    \xi(\phi) & = M_{Pl}^2 \kappa_1 e^{\kappa_2 \phi}\,, \\
    \phi & = \phi_0 \ln a^3(t)\,, 
\end{eqnarray}
where $\kappa_1, \kappa_2$ and $\phi_0$ are constants to be fixed with the only condition that the expected asymptotic form of the
function {$\delta$} is obtained. Similar considerations are made in the other three models.
Also, we assume that the solution for the metric resembles the
observed accelerated expansion, thus we take a constant Hubble factor, $H(z)=H_0$.
\item[Model B]Based on results   for cosmological scenarios\cite{Nojiri:2005vv}, we assume a power-law expansion, 
\begin{equation}
    \xi(\phi) =M_{Pl}^2  \kappa_1 e^{\kappa_2 \phi}\,, \ \ \ 
    \phi =  \phi_0 + \phi_1 \ln t\,, \ \ \ a(t) = a_0 t^x\,. \label{eq:modD}
\end{equation}
\item[Model C] A combination of the results for inflationary and cosmological scenarios ~\cite{Rizos:1993rt,Nojiri:2005vv}, both with a power law scale factor $a\sim t^x$ for some real exponent $x$, but with different forms of the potential and coupling function. Drawing from their results, we assume
\begin{equation}
    \xi(\phi) =M_{Pl}^2  \kappa_1 \phi^2\,, \ \ \ 
    \phi =  \phi_0 + \phi_1 t^2\,, \ \ \ a(t) = a_0 t^x\,.\label{eq:modC}
\end{equation}
\item[Model D] Based on a de Sitter solution in a model with quadratic coupling~\cite{Kanti:2015pda}. This solution is found in the study of early universe
dynamics; however, if the theory is to be compatible with observations, then at late times we expect the scale factor
to evolve as in a dark energy dominated universe, i.e., as a 
de Sitter solution. This motivates us to take the background 
solution
\begin{eqnarray}
    \xi(\phi) & =& M_{Pl}^2 \kappa_1 \phi^2\,, \\
    \phi & = &\phi_0 \exp\left( -\frac52 H_0 t \right)\,,
\end{eqnarray}
with a constant Hubble parameter $H_0$.
\end{description}
In all the equations above, $\kappa_1, \kappa_2, \phi_0, \phi_1$ and $a_0$ are constants and are specific for each model, i.e., a choice
of their values for Model A has no implication on their values for other models.
Models $A$, $B$, $C$ and $D$ are used as cosmological backgrounds where gravitational waves propagate. We emphasize that at this point we are interested in the fit between the
parametrizations and the theoretical predictions rather than in the actual physical viability of the models. To simplify the notation hereafter we set $M_{Pl}^2 = 1$.

\subsection{Modified friction term in EsGB gravity}
\begin{figure}
        \centering
        \includegraphics[width=0.47\textwidth]{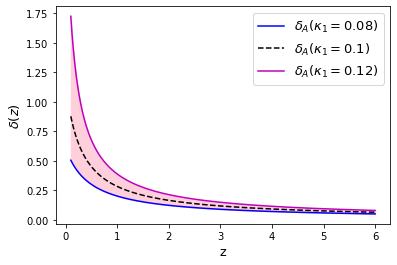} \includegraphics[width=0.47\textwidth]{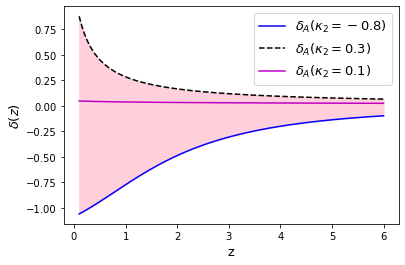}
        \caption{Profiles of $\delta(z)$ in model A for different choices of the parameter $\kappa_1$ (left) and $\kappa_2$ (right). The parameters that are fixed take the values given in Table~\ref{tab:models}. Similar results hold for the other three models.}
        \label{fig:modelachoices}
    \end{figure}

We take the variation  of  eq.~\eqref{eq:actionesgb} with respect to the metric $g^{\mu\nu}$ in order to obtain the field equations
of motion for EsGB gravity. Then we fix  $\xi(\phi)$ as either
a quadratic or exponential coupling, depending on the model we are considering, and we take first order tensor perturbations, $g_{\mu\nu} = \bar g_{\mu\nu}(x^\alpha)+h_{\mu\nu}(x^\alpha)$ with $|h_{\mu\nu}|\ll1$ 
around a background metric $\bar g_{\mu\nu}$ that we choose as 
a flat FRW metric, 
\begin{equation}
    d\bar s^2 = \bar g_{\mu\nu}d x^\mu dx^\nu = a^2(\eta) ( - d\eta^2 + d\Sigma^3 )\,,  
\end{equation}
where $d\Sigma^2$ is the line element of a flat, spacelike metric, and $\eta$ represents conformal time. Performing the perturbation of the equations of motion in traceless transverse gauge, neglecting corrections to the speed of light by the same arguments 
that lead to eq.~\eqref{eq:mglinpert}, and comparing with that same equation, we find that the $\delta$-modifications to the friction term for any $\phi$-dependent coupling are given by
\begin{equation}
\delta =    \frac{4 \left(\mathcal{H} \phi '^2 \xi_{\phi\phi}+\xi_\phi\left(\mathcal{H}'-2 \mathcal{H}^2\right) \phi '+\xi_\phi\mathcal{H} \phi ''\right)}{\mathcal{H} \left(  a^2-8   \mathcal{H} \phi ' \xi_\phi \right)}\,, \label{eq:deltaanyxi}
\end{equation}
where $\mathcal H = a'/a$, primes denote derivatives with 
respect to conformal time, $\xi$ is the coupling function and $\xi_\phi, \xi_{\phi\phi}$ are its first and second derivatives with respect to the scalar field. For the exponential (models $A$ and $B$) and quadratic (models $C$ and $D$) coupling functions, $\delta(z)$ takes the form
\begin{eqnarray}
    \delta_{A,B} & =&\frac{ 4 \kappa _1 \kappa _2   e^{\kappa _2 \phi }  \left[  \left(\mathcal{H}'-2 \mathcal{H}^2\right) \phi ' + \mathcal H \kappa_2 \phi'^2 + \mathcal H \phi'' \right] }
    {  \mathcal H [ a^2 - 8 e^{\kappa_2 \phi}  \mathcal H \kappa_1 \kappa_2 \phi' ]}\,,\\
\delta_{C,D} & =& \frac{8 \kappa _1    \left[ \left(\mathcal{H}'-2 \mathcal{H}^2\right)\phi \phi '  + \mathcal H \phi'^2 + \mathcal H \phi''  \right]}{\mathcal H[a^2 - 16  \phi \mathcal H \kappa_1 \phi']}\,.
\end{eqnarray}
After converting to cosmological time, using the background solution of each model, and expressing the results in terms of redshift, each $\delta(z)$ becomes
\begin{eqnarray}
    \delta_A(z) & =& \frac{36 \phi_0^2  H_0^2 \kappa _1 \kappa _2^2  }{(1+z)^{3 \phi_0 \kappa _2}-24 \phi_0 H_0^2 \kappa _1 \kappa _2  }\,, \\
    \delta_B(z) & =& \frac{4 \kappa_1 \kappa_2   \phi_1 e^{\kappa_2 \phi_0} (\kappa_2 \phi_1-2) ({1+z})^{-\kappa_2 \phi_1/x}}
    {({1+z})^{-2/x}-8 \kappa_1 \kappa_2   x \phi_1 e^{\kappa_2 \phi_0} \left({1+z}\right)^{-\kappa_2 \phi_1/x}}\,, \\
    \delta_C(z) & =& -\frac{32 \kappa _1   \phi _1^2 ({1+z})^{-2/x}}{32 \kappa _1   x \phi _1 \left(\phi _1 ({1+z})^{-2/x}+\phi _0\right)-1}\,,\\
    \delta_D(z) & =& \frac{100 \phi_0 ^2 H_0^2 \kappa _1  }{40 \phi_0 ^2 H_0^2 \kappa _1  +{(1+z)^{-5}}}.
\end{eqnarray}
In order for these functions to decay asymptotically, one must choose appropriate values for the constants involved. The values we have used are reported in Table~\ref{tab:models}. { These choices are rather arbitrary. Indeed,  Fig.~\ref{fig:modelachoices} shows that the qualitative behaviour of $\delta(z)$ (asymptotically vanishing) holds for several different choices of parameters in model A. Similar results hold for the other three models.}
For model $D$ it is impossible to get an asymptotically vanishing $\delta$, instead, it always approaches 
$\delta\to 5/2$. Thus, model $D$ modifies gravity at
every redshift and it should be ruled out by various 
observational tests. Furthermore, the parametrizations given by eq. \eqref{eq:param1}, \eqref{eq:param2} and \eqref{eq:param3} are specifically
designed to give $\delta\to0$ as $z\to\infty$. Nevertheless, we have decided to trivially modify these parametrizations by adding a constant term, so that we 
can retain model $D$ and gather more information on the
ability of  $\delta_{I},\delta_{II},\delta_{III}$ to reproduce the dynamical part of {$\delta$}. 

In principle one can use eqs.~\eqref{eq:lumdis} and~\eqref{eq:Xi} to write down the expression for the function $\Xi(z)$ corresponding to {$\delta$} for any coupling function. However, the
integral involved in such an expression is not always analytically solvable, and model $B$ provides an example of this. Therefore, for this case the fit is
performed between the parametrizations and a numerical profile. For the other three models, $\Xi(z)$ is computed analytically. The results are the following,
\begin{eqnarray}
   \Xi_A(z) &=& \frac{(1+z)^{\frac{3 \phi_0 \kappa_2}{2}} \sqrt{1-24 \phi_0 H_0^2   \kappa_1 \kappa_2}}{\sqrt{(1+z)^{3 \phi_0 \kappa_2}-24 \phi_0 H_0^2   \kappa_1 \kappa_2}},\\
    \Xi_B(z)& =&\exp\left( -\int_0^z \frac{\delta_D(z')}{z'+1} \, dz'\right) , \\
    \Xi_C(z) & = &\frac{\sqrt{32   \kappa_1 x \phi_1 (\phi_0+\phi_1)-1}}{\sqrt{32   \kappa_1 x \phi_1 \left(\phi_1 (1+z)^{-2/x}+\phi_0\right)-1}},
\\
    \Xi_D(z) &= &\frac{\sqrt{40 \phi_0 ^2 H_0^2   \kappa_1+1}}{\sqrt{40 \phi_0 ^2 {H_0}^2   \kappa_1 (1+z)^5+1}}\,.
\end{eqnarray}

In the next section we first present the results of fitting models $A$-$D$ with the parametrizations for $\delta(z)$ and $\Xi(z)$ given in eqs. \eqref{eq:param1} to \eqref{eq:paramxi3}, then we compare the best of these parametrizations to $\delta_X$ and $\Xi_X$.

\section{Confronting theory and parametrizations}\label{sec:results}
{ We present, in Fig.~\ref{fig:dhostdata}, an initial exploration of the viability of $\delta_X$ and $\Xi_X$  for fitting the predictions of a model of DHOST gravity. Now we want to see whether this parametrization also fits predictions of quadratic gravity, and if it does, we want to compare it to the fits provided by the parametrizations $I, II, III$. Thus, we begin this section by exploring these three parametrizations in quadratic gravity, showing explicitly, that $I$ works better than $II$ and $III$ for this class of models;  and then we compare the fits provided by the parametrizations $I$ and $X$. }

\begin{figure}
    \centering
    {\includegraphics[width=0.48\columnwidth]{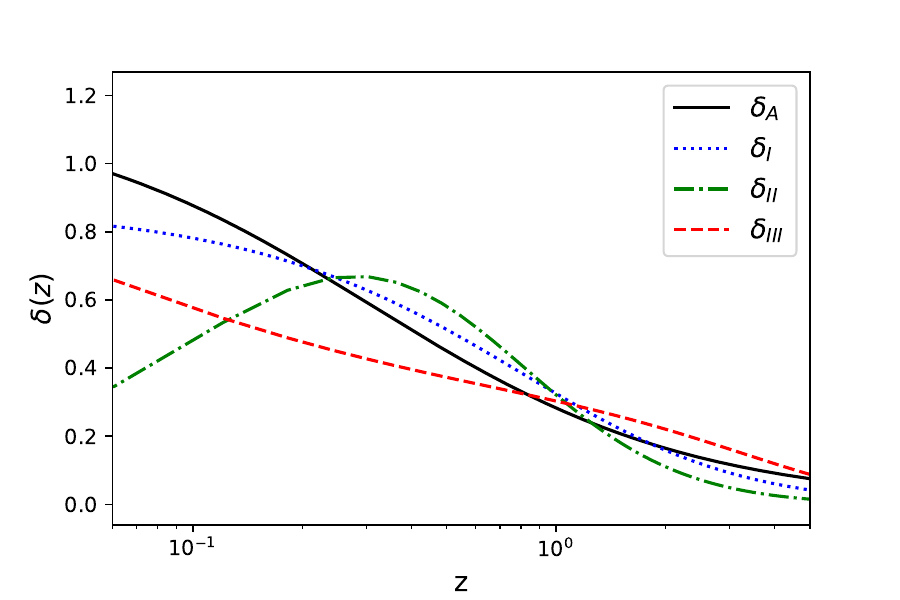}}
    {\includegraphics[width=0.48\columnwidth]{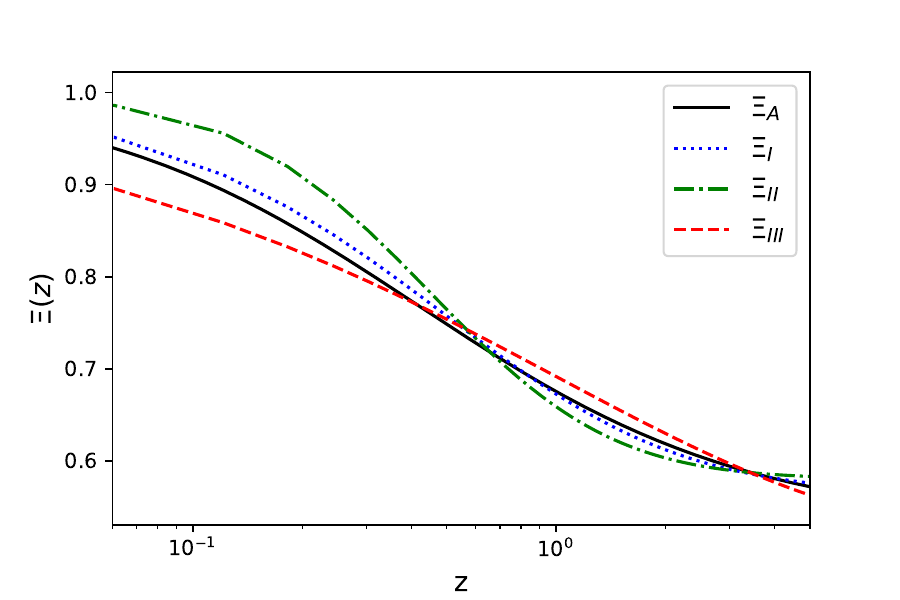}}
    {\includegraphics[width=0.48\columnwidth]{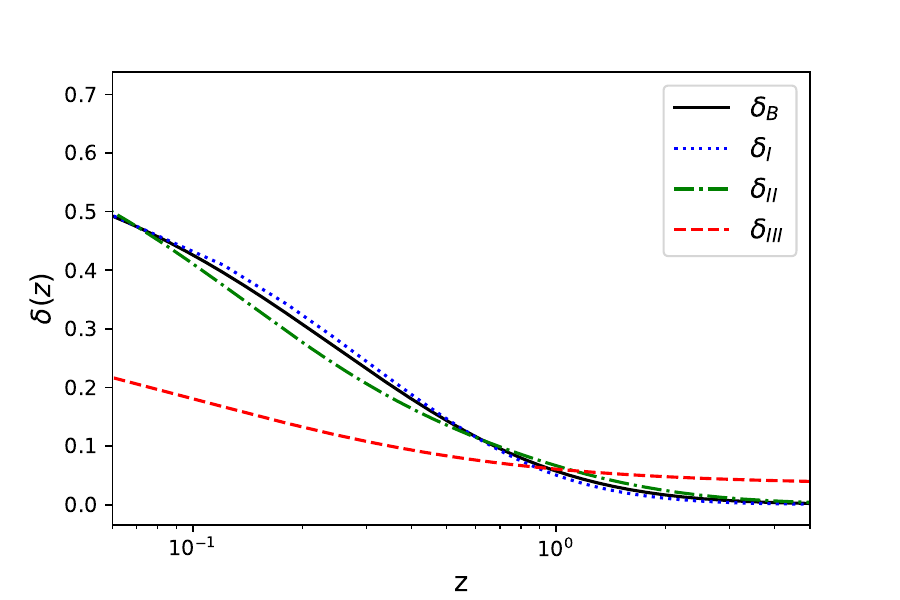}}
    {\includegraphics[width=0.48\columnwidth]{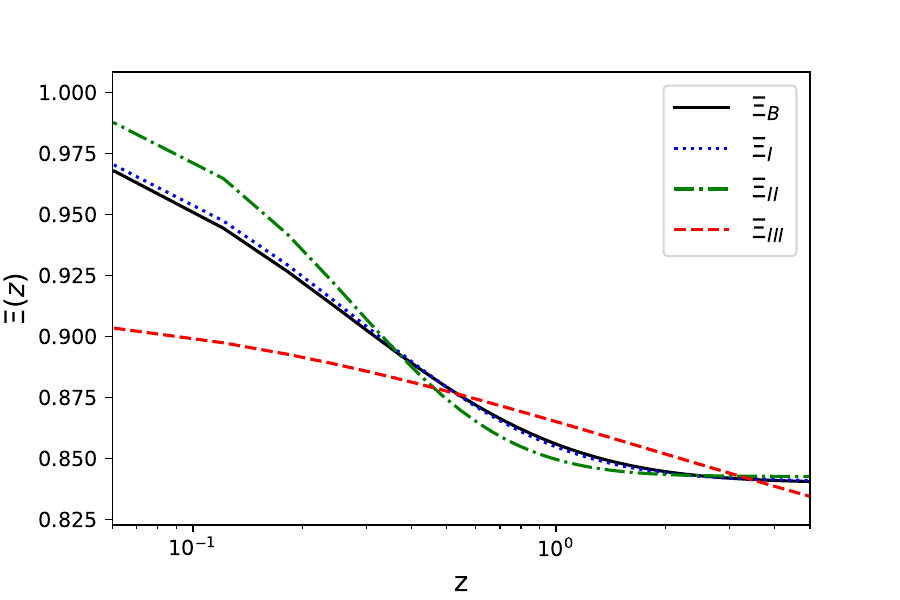}}
    {\includegraphics[width=0.48\columnwidth]{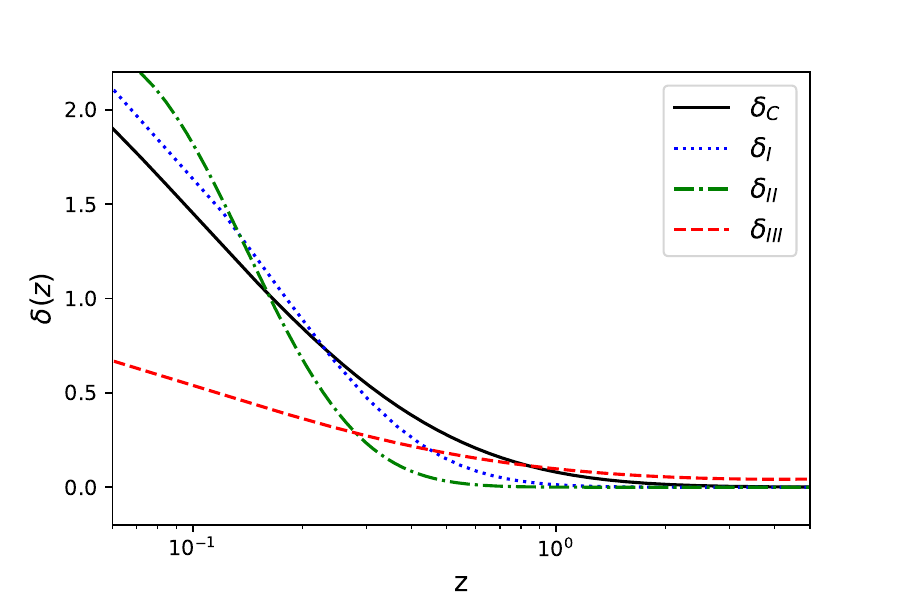}}
    {\includegraphics[width=0.48\columnwidth]{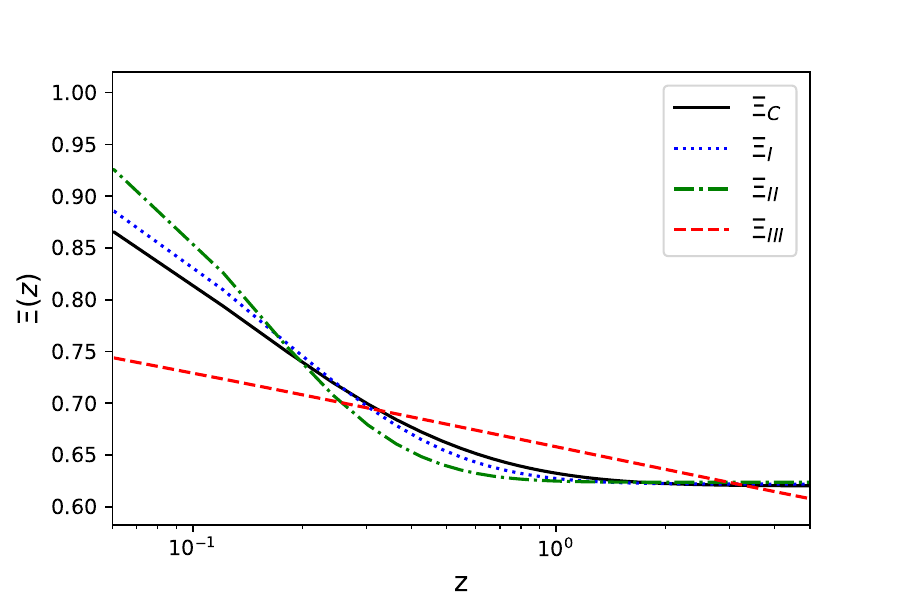}}
    {\includegraphics[width=0.48\columnwidth]{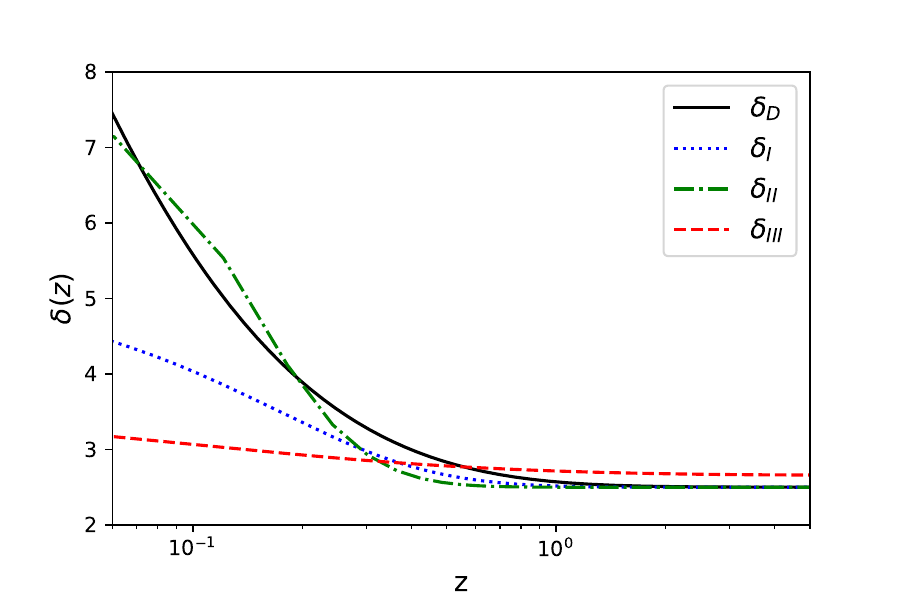}}
    {\includegraphics[width=0.48\columnwidth]{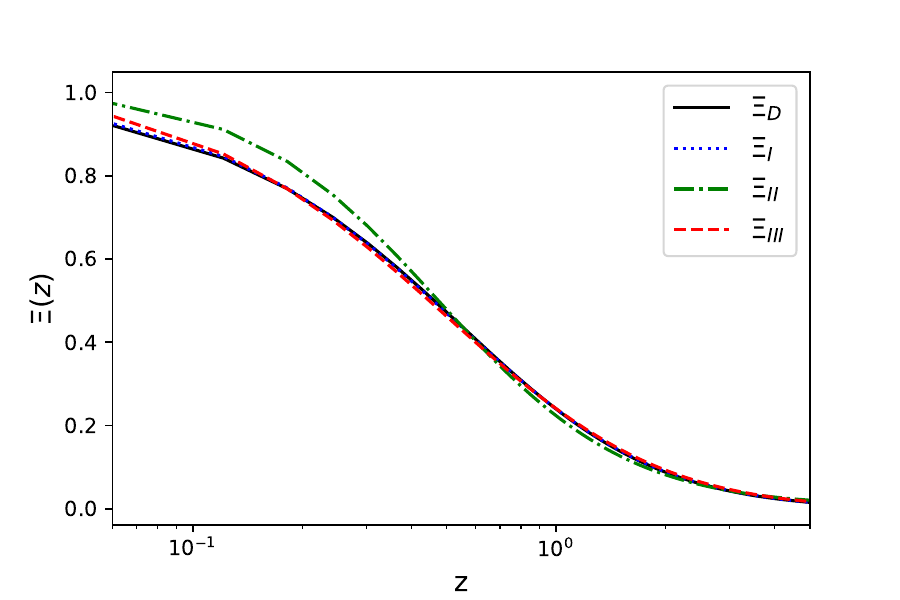}}
    \caption{Comparison between the functions $\delta$ and $\Xi$ for models $A$-$D$ and the parametrizations given by eqs.~\eqref{eq:param1}-\eqref{eq:paramxi3} with the values for the parameters reported in Table \ref{tab:paramvalues}. 
    }
    \label{fig:modcomparison}
\end{figure}

    In Fig.~\ref{fig:modcomparison} we show the theoretical profiles for {$\delta$}
    and $\Xi(z)$ for models A-D with the constants reported in Table~\ref{tab:models}, as well as the parametrizations $\delta_{I},\delta_{II},\delta_{III}$ (left panels) and $\Xi_{I},\Xi_{II},\Xi_{III}$ (right panels). By design
    $\delta_{II}(0) = 0$,  this is required in models where
    the cosmological solution evolves towards a de Sitter fixed point. Similarly, $\delta_{III}(0) = 0$ for $n>1$. This restriction makes $\delta_{II}$ and $\delta_{III}$ unable to
    reproduce the theoretical {$\delta$} in the models we 
    are studying when $z\to 0$. Since $\Xi(z)$ is a cumulative effect, it is  sensitive to the
    behaviour of {$\delta$} near $z\to 0$. Thus, if $\delta(z\sim 0)$ is not well fitted neither is $\Xi(z)$, unless we allow $\Xi(z)$ to select its best fit parameters independently of those of {$\delta$}. In this case, the parameters extracted from the fit to $\Xi(z)$ do not necessarily agree with those extracted from the fit to {$\delta$}. It is important to keep this in mind when fitting actual data: a good fit to $\Xi(z)$ does not imply that the same parameters provide a good fit for {$\delta$}, therefore, fitting one quantity is not enough to conclude that the parametrizations favour a certain model.

\begin{table}[h]
\tbl{Parameters for the models $A$-$D$ presented in Sec.~\ref{sec:EGB}. These parameters are chosen such that {$\delta$} vanishes asymptotically, except for model $D$ where this is not possible and we only require that $\delta'$ is asymptotically vanishing.  }
{\begin{tabular}{@{}llllll}
        \toprule    
        Model & $\kappa_1$ & $\kappa_2$ & $\phi_0$ & $\phi_1$ &  $x$ \\\colrule
        $A$ & 0.1 & 0.3 &  1.0 & - & - \\
        $B$ & 0.1 & -1.0 &  1.0 & 1.0 &  -1.0  \\
         $C$ & 0.1 &  - & -1.0 & 1.0 & 0.5 \\
         $D$ & -0.05 &  - & 1.0 & -& - \\ \botrule
    \end{tabular}\label{tab:models}}
\end{table}
\begin{table}[h]
\tbl{Parameters $(n_0,\Xi_0)$ determined by a least squares fit between the parametrizations I-III and the models $A$-$D$ specified in Table~\eqref{tab:models}. The fits to $\delta$ and $\Xi$ are performed independently, therefore the values of $n_0$ and $\Xi_0$ obtained by each fit are not necessarily self-consistent.}
{\begin{tabular}{@{}ccccc}
        \toprule
        & A & B & C & D \\ \colrule
        \multicolumn{5}{c}{$\delta_I$}\\
        \hline
        $n_0$ & $2.067\pm0.024$ & $3.79\pm0.016$ & $8.43\pm0.09$ & $8.589\pm0.023$\\ 
        $\Xi_0$ & $0.561\pm0.003$ & $0.843\pm0.001$ & $0.65\pm0.003$ & $0.411\pm0.005$\\ \colrule
        \multicolumn{5}{c}{$\Xi_I$}\\
        \colrule
        $n_0$ & $2.006\pm0.011$ & $3.5\pm0.0075$ & $6.135\pm0.036$ & $3.37\pm0.017$ \\ 
        $\Xi_0$ & $0.56\pm0.0006$ & $0.84\pm0.00005$ & $0.62\pm0.0002$ & $1.4\pm0.001$\\ \colrule
        \multicolumn{5}{c}{$\delta_{II}$}\\
        \colrule
        $n_0$ & $2.908\pm0.099$ & $6.16\pm0.268$ & $9.69\pm0.7$ & $14.63\pm3.2$\\ 
        $\Xi_0$ & $0.626\pm0.01$ & $0.87\pm0.005$ & $0.7\pm0.017$ & $0.514\pm0.06$\\ \colrule
        \multicolumn{5}{c}{$\Xi_{II}$}\\
        \colrule
        $n_0$ & $3.07\pm0.038$ & $5.36\pm0.055$ & $9.04\pm0.12$ & $2.42\pm0.012$ \\ 
        $\Xi_0$ & $0.58\pm0.0012$ & $0.84\pm0.0002$ & $0.624\pm0.0005$ & $0.008\pm0.0011$\\ \colrule
        \multicolumn{5}{c}{$\delta_{III}$}\\
        \colrule
        $n_0$ & $0.695\pm0.007$ & $0.001\pm12.8$ & $0.001\pm1.9$ & $0.0\pm7.662$\\ 
        $\Xi_0$ & $0.568\pm0.004$ & $-0.56\pm365.2$ & $-0.58\pm11.6$ & $-0.58\pm63.26$\\ \colrule
        \multicolumn{5}{c}{$\Xi_{III}$}\\
        \colrule
        $n_0$ & $0.5\pm0.0063$ & $0.26\pm0.008$ & $0.158\pm0.006$ & $0.62\pm0.008$ \\
        $\Xi_0$ & $0.512\pm0.0012$ & $0.79\pm0.001$ & $0.46\pm0.0021$ & $1.57\pm0.004$ \\ \botrule
    \end{tabular}\label{tab:paramvalues}}
\end{table}
    In Table~\ref{tab:paramvalues} we report the best fit parameters obtained from independent fits for $\delta(z)$
    and $\Xi(z)$ with the three parametrizations described in 
    Sec.~\ref{sec:linwave}. Either from Fig.~\ref{fig:modcomparison} or from the errors reported in Table~\ref{tab:paramvalues} we conclude that the parametrization $\Xi_I$ performs better than the others. A few remarks about these figures are in order.

    \begin{itemize}
        \item  Model $D$ (fourth row in Fig.~\ref{fig:modcomparison}) modifies gravity at all redshifts. Asymptotically, its {$\delta$} approaches a constant value of $2/5$ which is not compatible with the parametrizations given by eq. \eqref{eq:param1}, \eqref{eq:param2} and \eqref{eq:param3}. Thus, only in this case we minimally modified these parametrizations by the addition of a constant term. At this point, one could either fix this constant term to the known asymptotic value of {$\delta$} or leave it as another free parameter to be fixed by the least squares methods. 
        We took the former approach.
        In addition, for this model we modified $\Xi_{I}$, $\Xi_{II}$, $\Xi_{III}$
        in order to take into account the constant asymptotic value of {$\delta$}.
        \item Model $D$ also exemplifies that $\Xi(z)\to0$ as $z\to\infty$ is as an indication that GR is being modified at all redshifts. If the integral in~\eqref{eq:lumdis} grows for large $z$, which happens for instance if {$\delta$} is constant, then $\Xi(z)$ goes to zero. 
        \item The best fit parameters obtained with $\delta_{III}$ display large uncertainties. These uncertainties can be reduced by removing from the analysis points that are too close to the origin, i.e., by changing the range of $z$ used in the numerical fits. However, in order to report parameters obtained under the same search criteria, we did not do this modification here.
    \end{itemize}
    
The results above show that $\delta_I$ and $\Xi_I$ provide good fits to well motivated quadratic models of gravity, in particular to models $A$ and $B$ that are motivated by late time cosmological solutions.
Nevertheless, as we mentioned before these parametrizations are not suitable for certain types of modifications of gravity, in particular {some examples} of DHOST gravity. 
On the other hand, $\delta_X$ offers a good fit both for the models studied in Sec.~\ref{sec:dhost} and for the EsGB models, as we discuss below. {Also, in Fig.~\ref{fig:modcomparison} we notice that the models $C$ and $D$ are similar to $A$ and $B$ regarding the shapes of $\delta$ and $\Xi$, thus, in the next discussions we focus on models $A$ and $B$.}
 


\begin{table}
        \tbl{The first two columns show the parameters $(\Xi_0,n,r)$ determined by least squares fits between the parametrizations $\delta_X$, $\Xi_X$ and the models $A$ and $B$ specified in Table~\ref{tab:models}. The third column shows the parameters obtained by fitting $\delta_X, \Xi_X$ to their predictions for GW propagation in DHOST gravity. 
        The fits to $\delta$ and $\Xi$ are performed independently, therefore the values of $n$ and $\Xi_0$ obtained by each fit are not necessarily self-consistent.}
{\begin{tabular}{@{}cccc}
        & A & B & DHOST \\ \colrule
        \multicolumn{4}{c}{$\delta_X$}\\
        \colrule
        $\Xi_0$ & $0.69\pm0.0014$ & $0.9\pm0.0003$ & $1.02\pm0.0001$ \\ 
        $n$ & $1.687\pm0.013$ & $3.637\pm0.013$ & $2.823\pm0.017$\\ 
        $r$ & $1.743\pm0.013$ & $3.831\pm0.11$ & $2.612\pm0.012$\\ \colrule
        \multicolumn{4}{c}{$\Xi_X$}\\
        \colrule
        $\Xi_0$ & $0.66\pm0.0006$ & $0.89\pm0.0001$ & $1.017\pm0.0001$\\ 
        $n$ & $2.084\pm0.012$ & $3.887\pm0.012$ & $2.247\pm0.024$\\ 
        $r$ & $2.085\pm0.012$ & $3.888\pm0.012$ & $2.247\pm0.024$ \\ \botrule
\end{tabular}
\label{tab:paramvalues2}}
\end{table}

\begin{figure}
      \includegraphics[width=0.49\textwidth]{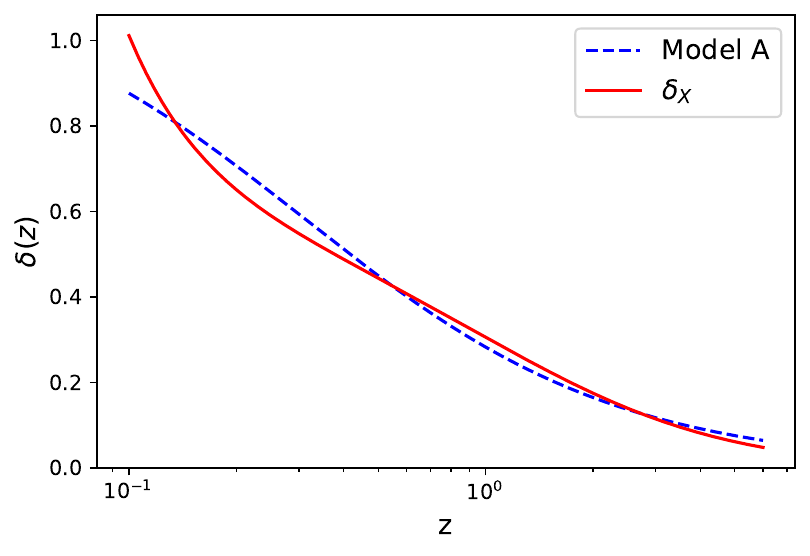}
      \includegraphics[width=0.49\textwidth]{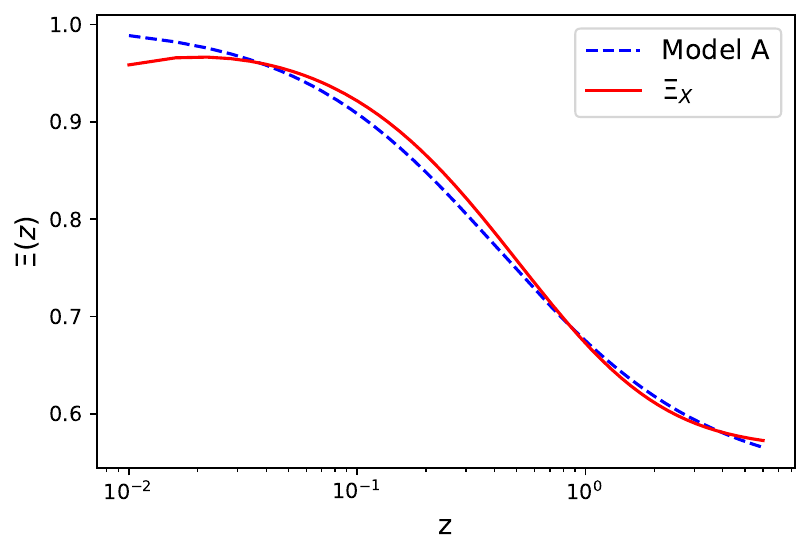} \\
      \includegraphics[width=0.49\textwidth]{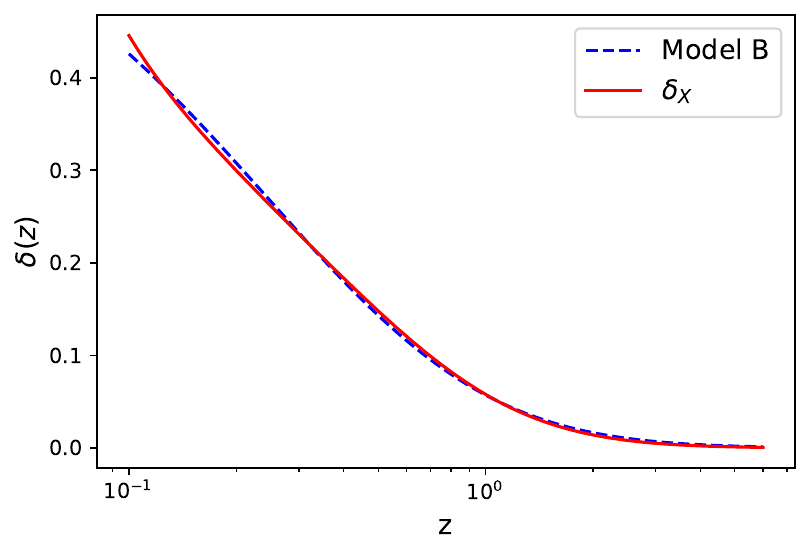}
      \includegraphics[width=0.49\textwidth]{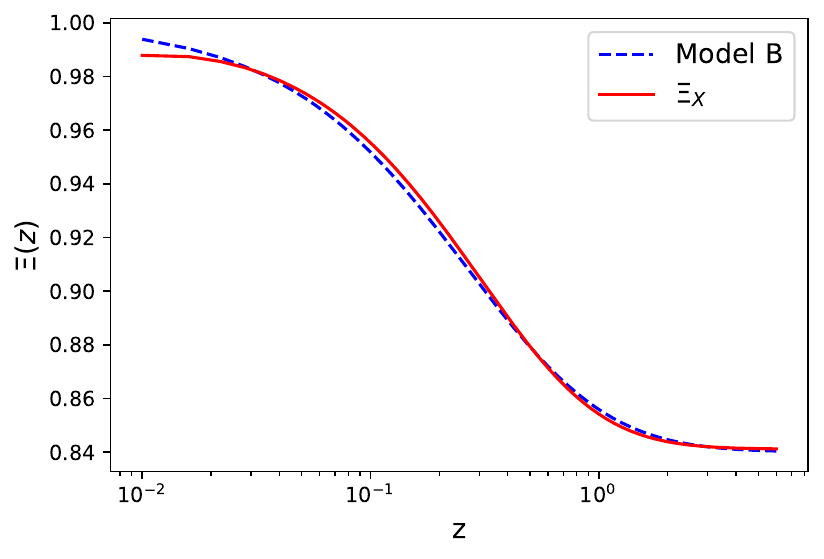}
      \caption{Parametrizations $\delta_X$ and $\Xi_X$ applied to models $A$ (de Sitter expansion, top panels) and $B$ (power law expansion, bottom panels). The best fit parameters are reported in Table~\ref{tab:paramvalues2}. 
      }
      \label{fig:dX}
\end{figure}

In the top panels of Fig.~\ref{fig:dX} we show the results of a least squares fit between $\delta_X,\Xi_X$ and the predictions of  $A$, $B$ and DHOST models. The best fit parameters are reported in the first and second columns of Table~\ref{tab:paramvalues2}. Comparing with Table~\ref{tab:paramvalues}, we see that the fit is similar to the one obtained with $\delta_I$ and $\Xi_I$. This is confirmed by the {estimators} reported in Table~\ref{tab:chi} which show a significant improvement for the fits to {$\delta$}.  
\begin{table}
\tbl{Statistical estimators for the fits to Models A and B with the parametrizations $\delta_X$, $\Xi_X$ compared to the best of the paramerizations $I, II$ and $III$, i.e., to parametrization $I$. The fit for $\delta(z)$ in model $A$ presents a significant improvement when using $\delta_X$, while in all the other cases the goodness of fit is similar. }
{\begin{tabular}{@{}ccccc}
       & \multicolumn{4}{c}{Model A} \\ 
\cline{1-5} 
        & $\delta_I$ & $\delta_X$  & $\Xi_I$ & $\Xi_X$\\ \colrule
        $\chi^2$ & $11.42$ & $4.33$ & $0.442$ &  $0.472$\\ \botrule
        $\chi^2_{red}$ & $0.1165$ & $0.0446$ & $0.0045$ & $0.048$ \\ \colrule
        $\mathcal{R}^2$ & $0.9624$ & $0.9857$ & $0.9946$ & $0.9943$ \\ \botrule
        $\mathcal{R}^2_{adj}$ & $0.9255$ & $0.9711$ & $0.9892$ &  $0.9884$\\ \botrule
    \end{tabular}
   \begin{tabular}{@{}ccccc}
       & \multicolumn{4}{c}{Model B} \\ 
\cline{1-5} 
        & $\delta_I$ & $\delta_X$  & $\Xi_I$ & $\Xi_X$ \\ \colrule
        $\chi^2$ & $0.112$ & $0.075$ & $0.009$ & $0.017$ \\ \botrule
        $\chi^2_{red}$ & $0.0011$ & $0.000773$ & $0.0000918$ & $0.000175$ \\ \colrule
        $\mathcal{R}^2$ & $0.9982$ & $0.9988$ & $0.9989$ & $0.9980$ \\ \botrule
        $\mathcal{R}^2_{adj}$ & $0.9965$ & $0.9976$ & $0.9978$ &  $0.9959$\\ \botrule
    \end{tabular}\label{tab:chi}}
\end{table}

From the results of the previous section, we also conclude that $\delta_X$ and $\Xi_X$ perform better than the parametrizations labeled $II$ and $III$. In order to summarise the results for $\delta_X$ and $\Xi_X$, Table~\ref{tab:paramvalues2} also shows the best fit parameters for DHOST gravity. Notice that for DHOST gravity the parameters obtained from $\delta_X$ satisfy $\Xi_0>1$ and $n>r$, while for EsGB we have $\Xi_0<1$ and $n<r$. These properties are related to the profile of {$\delta$} near the origin.


\section{Discussion}\label{sec:disc}

In this work we analyse three existing parametrizations for the propagation of gravitational waves. In particular, we study how these 
parametrizations fit the predictions of different cosmological models in EsGB and DHOST gravity, including models that describe late time acceleration of the universe and that are able to fulfil the stringent constraint on the speed of gravitational waves imposed by the observation of GW170817. We focus on the robustness of the parametrizations rather than on the viability of the models we use. We find that the simplest of these parametrizations gives good agreement with the predictions of
EsGB. The other two parametrizations, labelled $II$ and $III$,  are designed for models where {$\delta$} goes to zero at late times, for instance due to the presence of a de Sitter fixed point and a tracker cosmological solution, as in the DHOST model analysed in Sec.~\ref{sec:dhost}. This characteristic is absent in the EsGB models that we studied, therefore their predictions for {$\delta$} are not well fitted by $\delta_{II}$ and $\delta_{III}$. On the other hand, even for these models the parametrization $\Xi_{II}$ gives a good approximation to the predicted values of $\Xi(z)$. This emphasises the robustness of  $\Xi(z)$; since it is an integrated effect it is less sensitive to the local modifications to the friction term in the gravitational wave equation. From these results we conclude that both $\Xi_I$ and $\Xi_{II}$
are viable parametrizations for quadratic gravity.

Due to their simplicity, parametrizations $I$, $II$ and $III$ are unable to fit the diverse predictions of different models of MG. 
Since in this work our focus is on the generality of the parametrizations, we propose a novel and more general profile, $\delta_X$, that contains an additional parameter  allowing the fit to select between models that behave like  $\delta_I$ and models that behave like $\delta_{II}$. In addition to the generality of $\delta_X$, the values of the $\chi^2$ test show a significant improvement with respect to $\delta_I$, both for EsGB and DHOST theories. {On the other hand, for
$\Xi(z)$ a fit with only two parameters suffices, as can be concluded from comparing $\Xi_I$ and $\Xi_X$ in Table~\ref{tab:chi} and Figs. \ref{fig:modcomparison} and \ref{fig:dX}.}

Once the parameters $n,r$ and $\Xi_0$ are determined by comparing to actual data, one should try to identify a physical model that is well described by these parameters. 
Along these lines, our results show that the best fit parameters obtained independently from {$\delta$} and $\Xi(z)$ do not necessarily agree, therefore model identification based only on one of those functions might be inconsistent. It is natural to expect that this is alleviated either by incorporating other sets of data in addition to gravitational waves or by performing a simultaneous optimisation of {$\delta$} and $\Xi(z)$.

\subsection*{Acknowledgments}
CO and JC acknowledge the support provided by UAZ-2018-37554. AL is supported by CONACyT graduate grant No. 855158. 
\vspace{2cm}
\bibliographystyle{unsrt}
\bibliography{ref}
\end{document}